\documentclass[conference]{IEEEtran}
\IEEEoverridecommandlockouts

\usepackage{cite}
\usepackage{amsmath,amssymb,amsfonts}
\usepackage{algorithmic}
\usepackage{graphicx}
\usepackage{textcomp}
\usepackage{xcolor}
\usepackage{float}
\usepackage{booktabs}

\def\BibTeX{{\rm B\kern-.05em{\sc i\kern-.025em b}\kern-.08em
    T\kern-.1667em\lower.7ex\hbox{E}\kern-.125emX}}
\begin{document}

\title{Digit Recognition using Multimodal Spiking Neural Networks}

\author{William Bjorndahl*, Jack Easton*, Austin Modoff, Eric C. Larson, Joseph Camp, Prasanna Rangarajan\\
\text{Southern Methodist University}\\
\thanks{* These authors are equal contributors.}
}

\maketitle

\begin{abstract}
Spiking neural networks (SNNs) are the third generation of neural networks that are biologically inspired to process data in a fashion that emulates the exchange of signals in the brain. Within the Computer Vision community SNNs have garnered significant attention due in large part to the availability of event-based sensors that produce a spatially resolved spike train in response to changes in scene radiance. SNNs are used to process event-based data due to their neuromorphic nature. The proposed work examines the neuromorphic advantage of fusing multiple sensory inputs in classification tasks. Specifically we study the performance of a SNN in digit classification by passing in a visual modality branch (Neuromorphic-MNIST [N-MNIST]) and an auditory modality branch (Spiking Heidelberg Digits [SHD]) from datasets that were created using event-based sensors to generate a series of time-dependent events. It is observed that multi-modal SNNs outperform unimodal visual and unimodal auditory SNNs. Furthermore, it is observed that the process of sensory fusion is insensitive to the depth at which the visual and auditory branches are combined. This work achieves a 98.43\% accuracy on the combined N-MNIST and SHD dataset using a multimodal SNN that concatenates the visual and auditory branches at a late depth.
\end{abstract}


\section{Introduction}
\label{sec:introduction}

The need for more efficient and accurate models in fields such as robotics, autonomous vehicles and drones, and multimedia processing has driven the development of Artificial Neural Networks (ANNs). A common criticism levied against the current crop of ANNs is their rudimentary emulation of brain functionality, which has remained the principal motivation for the development of ANNs \cite{mazzoni1991more}. As an example, consider the case of signal propagation through an ANN, which is often represented as a matrix product followed by a non-linear compounding and thresholding. Though this structure is inspired by the functionality of the brain, the model drastically simplifies information propagation in synaptic connections \cite{eluyode2013comparative}. The brain of a living being works by propagating electrical signals, or action potentials, across neurons through a network of synaptic connections. Modeling these biologically plausible synaptic connections has motivated the development of SNNs \cite{ghosh2009spiking,tavanaei2019deep}. 

Previous research has focused on developing the biologically inspired functionality of SNNs. Unlike traditional ANNs, SNNs model the synaptic connections as time-dependent processes, capturing the rising and falling dynamics of neuronal action potentials. This time series modeling makes SNNs more suitable for tasks that involve temporal dependencies, such as speech recognition, vision analysis, and multisensory integration. Despite these advancements, the majority of studies have not fully explored the potential of SNNs in the context of multimodal data fusion. Existing studies such as reported in \cite{10.3389/fnsys.2022.845177}, have demonstrated the capability of SNNs to integrate multiple sensory modalities. However, the performance of multimodal integration against individual sensory modalities was not thoroughly compared. While recent advancements in neuromorphic sensors and computation devices have significantly improved the applicability of SNNs, there remains a gap in understanding how different sensory inputs can be effectively combined within these networks.

In this study, we aim to address the impact of incorporating both auditory and visual information into SNNs on their ability to classify digits from zero through nine. By comparing the performance of SNNs trained with combined auditory-visual data against those trained with individual modalities, we seek to demonstrate the advantages of multimodal integration in enhancing the accuracy and robustness of SNN models. Furthermore, we will explore performance differences of combining the auditory and visual modalities at different depths of the network. Our findings will contribute to the ongoing development of more biologically plausible and computationally efficient neural networks, with potential applications in areas that require precise and reliable sensory processing.

\section{Related Works}
\label{sec:relatedworks}
\begin{figure}[t]
\centering
\begin{tabular}{cc}
\includegraphics[scale=0.3]{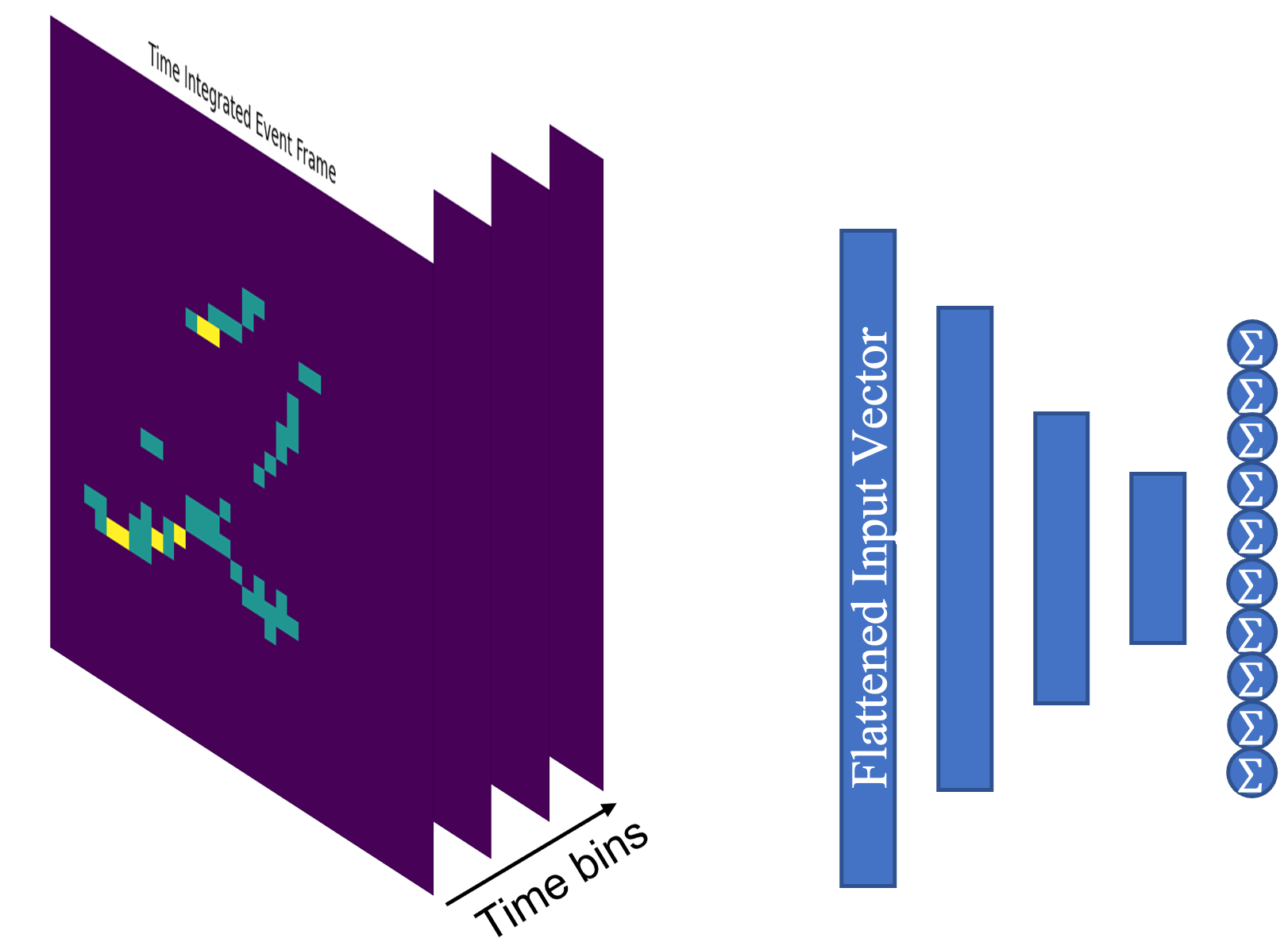} &
\includegraphics[scale=0.3]{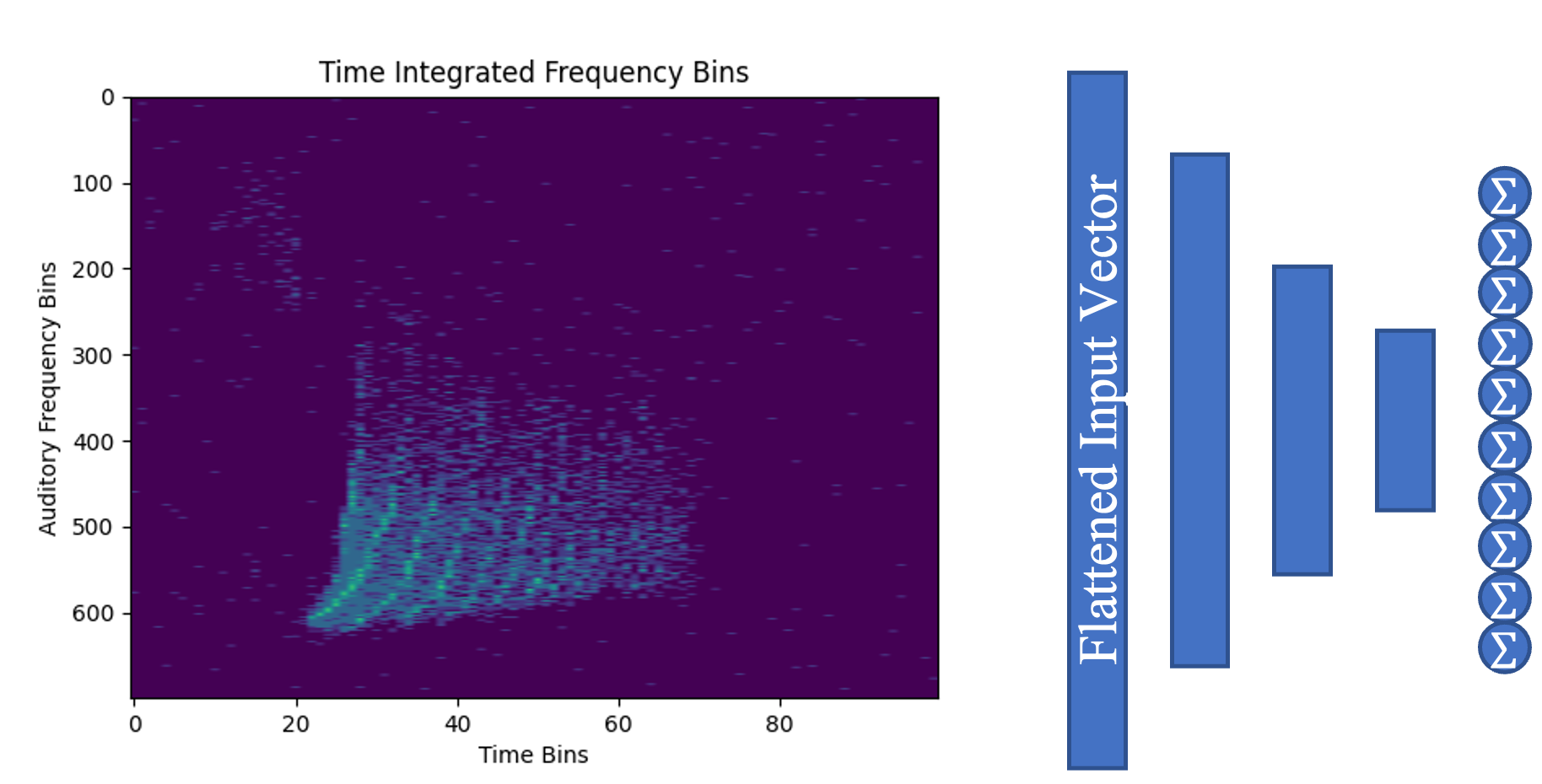} \\
\small{a) Visual input (N-MNIST)} & \small{b) Auditory input (SHD)}
\end{tabular}
\caption{Unimodal Network structures showing example data (digit two) as input for (a) visual and (b) auditory modalities.}
\label{fig:unimodal_diagram}
\end{figure}

While most work investigating SNNs has focused on singular modalities, such as digit recognition using either auditory or visual datasets (Fig.~\ref{fig:unimodal_diagram}), fewer studies have explored multimodal SNN models. The exploration of multimodal models is particularly relevant when dealing with noisy or incomplete data in one modality, where cross-modal integration can enhance the robustness and overall performance of the system. For example, \cite{9746865} investigated an attention-based cross-modal subnetwork that assigns attention scores in both auditory and visual branches. These scores are adjusted based on the quality of the input, with the branches being concatenated right before classification. This approach is valuable for situations where one of the input modalities is noisy, ensuring the model can still perform effectively.

Unimodal SNNs have achieved high accuracy in various digit recognition tasks. For instance, spike-based back propagation \cite{10.3389/fnins.2018.00331} and spike time dependent plasticity (STDP) \cite{Kheradpisheh_2018} in SNNs both achieved a high accuracy on the MNIST dataset. These results demonstrate the effectiveness of different learning mechanisms within SNNs, though they primarily focus on achieving high accuracy within a single modality.

Multimodal SNNs have been explored to a lesser extent, but they offer potential benefits in integrating complementary information from different sensory modalities. For example, \cite{MIDDYA2022108580} fused auditory and visual modalities at various levels of a convolutional neural network for multimodal emotion recognition, revealing that the level at which these features are combined can significantly affect the model’s performance. Our work expands on this idea by proposing three multimodal SNN architectures that extract event-based features at early, middle, and late levels for digit recognition (Fig.\ref{fig:multimodal_diagram}). In Section~\ref{sec:results}, we demonstrate that our multimodal SNN outperforms unimodal SNNs in digit classification across all levels of modality combination. Additionally, we show through the McNemar statistical test that the performance of our multimodal architecture is robust, regardless of the depth at which the modalities are integrated. These findings suggest that multimodal SNNs not only enhance accuracy but also contribute to a more flexible and resilient model architecture, making them advantageous for complex sensory processing tasks.

\section{Methods}
\label{sec:methods}

\subsection{Datasets}

The N-MNIST dataset \cite{10.3389/fnins.2015.00437} was created using a 34x34 pixel event-based visual sensor that was tilted and panned to generate asynchronous events from static MNIST data projected on a monitor, resembling the retinal movements observed in primates and humans when performing recognition tasks. We integrated the raw events in time bins to generate sparse frames as input to the visual networks. Each time bin is roughly 3 milliseconds long (100 time bins per instance).

The SHD dataset \cite{cramer_heidelberg_2020} contains spoken digits from 0 to 9 in both English and German for a total of twenty output classes. The audio was recorded in studio and converted into spiking events. We utilized the ten English classes for the auditory model prediction task. Similar to N-MNIST, we integrated the raw events into distinct time bins so that we could pass the sparse event frames into our network. Each time bin is roughly 7 milliseconds long (100 time bins per instance).

 Each pair of auditory and visual instances were grouped and aligned according to their output classes. We were limited by the size of the SHD dataset, because of this paired nature. Since SNNs are time-dependent, the number of time bins are kept equal between the auditory and visual datasets (100 total time bins for each instance modality).

\subsection{Network Characteristics}

\begin{figure}[t]
\centering
\includegraphics[scale=.3]{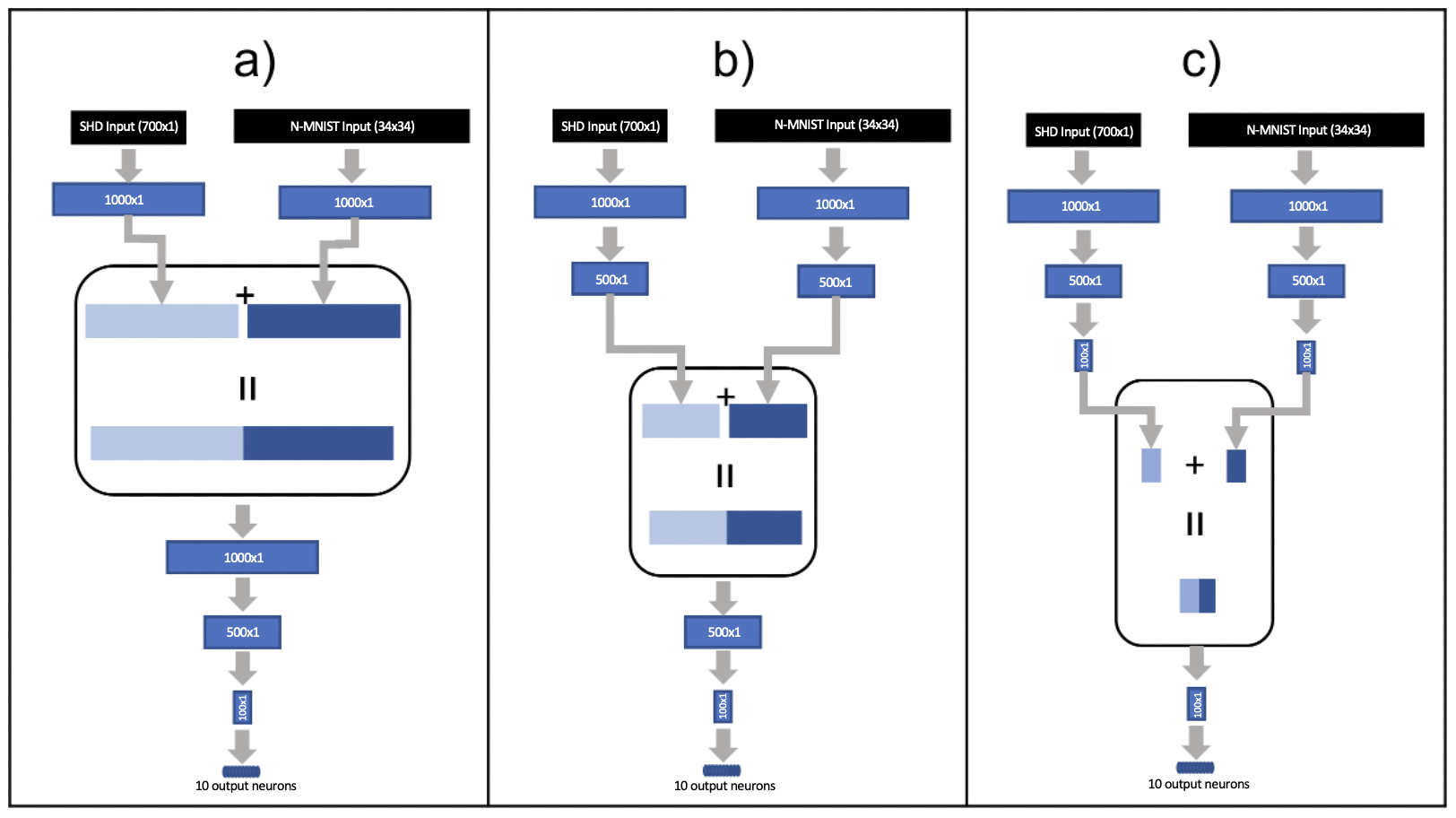}
\caption{Comparison of combining the visual and auditory branches at a (a) early, (b) middle, and (c) late depth in our multimodal SNN architecture.}
\label{fig:multimodal_diagram}
\end{figure}

The foundational unit of our SNN is the leaky integrate and fire (LIF) spiking neuron. This is the simplest and most common model of a biological neuron used in SNNs. The continuous time dynamics of LIF neurons are defined by their membrane potential \(v(t)\), which changes in response to the supplied input current \(I(t)\) according to equation (\ref{eq:1}):

\begin{equation} \label{eq:1}
\tau_v \cdot \frac{\partial}{\partial t}v(t) = -(v(t) - v_{rest} + R\cdot I(t),
\end{equation}

\begin{table*}[htbp]
\centering
\caption{Test accuracy comparison of our work with unimodal and multimodal methods. * represents works that use a multimodal SNN.}
\begin{tabular}{lccc}
{\textbf{Method}} & {\textbf{Dataset}} & {\textbf{Accuracy (\%)}} \\
\toprule
RSNN w/ Adaptation \cite{sparch}                 &     SHD & 94.6        \\
Feed-forward SNN with STFs \& attention \cite{10.3389/fnins.2022.1079357}              &     SHD & 92.4        \\

Unsupervised STDP \cite{7966074}                & N-MNIST          & 80.63                        \\

Back Propagation \cite{10.3389/fnins.2016.00508}             & N-MNIST          & 98.66                        \\

Attention Mechanism* \cite{9746865}                               &      MNIST-DVS + TIDIGITS                  &                98.95              \\
\hline
This work: Visual Only      & N-MNIST                & 92.25                        \\
This work: Auditory Only       & SHD                    & 95.29                        \\
This work: Early Fusion*    & N-MNIST + SHD          & 97.35                        \\
This work: Middle Fusion*   & N-MNIST + SHD          & 97.45                        \\
This work: Late Fusion*     & N-MNIST + SHD          & 98.43                        \\

\end{tabular}
\label{tab:accuracies}
\end{table*}
Where \(v_{rest}\) is a resting value \(v(t)\) exponentially decays to, \(R\) is the membrane resistance and \(\tau_v\) represents the time constant for the exponential decay of the voltage potential in the absence of an input current. When there is an input current, the neuron integrates the stimuli and adjusts the membrane potential accordingly. 
When the membrane potential \(v(t)\) surpasses the predefined threshold $v_{th}$, a spike is emitted and the potential is reset to \(v_{rest}\).

\subsubsection{Discrete Time Conversion}
On a piece of neuromorphic hardware, we could run events through our SNN in real-time and directly from our input sensors, which would more closely follow the above continuous-time equations for our neuron dynamics. Because of limitations from classical computing devices, GPUs and CPUs, we use discrete approximations of the continuous-time equations for the LIF neuron.

Similar to \cite{sparch}, we decided to use the forward-Euler first-order exponential integrator method with a step size of \(\Delta t = 1 ms\) to accomplish this conversion. The necessary continuous variable substitutions presented in \cite{sparch} are as follows,

\begin{equation} \label{eq:2}
v \rightarrow \frac{v - v_{rest}}{v_{th} - v_{rest}}, \\
I \rightarrow \frac{RI}{v_{th} - v_{rest}}
\end{equation}
The forward pass through our neuron thus becomes defined by the following discrete time equations:

\begin{gather} \label{eq:3}
    v[t] = \alpha\cdot v[t-1] + (1 - \alpha)\cdot I[t] - v_{th} \cdot s[t-1] \\
    s[t] = v[t] > v_{th}
\end{gather}
Intuitively, the first term in equation \eqref{eq:3} refers to the membrane potential leak, the second term is the excitation, and the third represents the effect of having spiked at the previous time step. The neuronal parameters in the above equations have been redefined as

\begin{equation}
\begin{gathered}
    v_{th} \rightarrow \frac{v_{th}-v_{rest}}{v_{th}-v_{rest}} = 1, \\
    \alpha = \exp(- \Delta t / \tau_v) \in [0.60, 0.96]
\end{gathered}
\end{equation}
All ranges are based on physiologically plausible values. 

\subsubsection{Network Readout and Loss Calculation}
At the output layer, we no longer want to output discrete-time spikes. Instead, we want predictive probabilities at each of the ten output neurons. Each of these neurons corresponds to the class of the output digit that is used during final prediction. This is similar to how many ANNs make predictions on their output neurons, with the addition of a \(softmax\) function.

We used the method described by \cite{sparch} to convert discrete-time spiking data to a single prediction probability \(P_i\) at each neuron (\(i = 1, 2, ..., N\)) in the final layer \(L\). Once the input sequence (of length \(T\) time steps) has been run fully through the network, we make a prediction using a cumulative sum of membrane potentials for each output neuron over time. The dynamics of this summation neuron are as follows,

\begin{equation}
P_i = \sum_{t=0}^T\text{softmax}(v_i^{(L)}[t]).
\end{equation}

The neuron that produces the highest cumulative sum over the full time \(T\) is selected as the prediction for the corresponding input. Then cross-entropy loss between the readout and the true values for the data is used to calculate the error.

\subsubsection{Surrogate Gradient Methods}
Each LIF neuron is effectively represented by a step function located at the voltage threshold. This function is not differentiable, therefore during backward propagation a differentiable surrogate step function is utilized for gradient calculation at each neuron. We utilize the boxcar surrogate function defined by \cite{kaiser} due to its low computational expense.

\begin{equation}
    \frac{\partial s[t]}{\partial v[t]} = \begin{cases}
        0.5 & |v[t] - v_{th}| \leq 0.5 \\
        0 & \text{otherwise}
    \end{cases}
\end{equation}

\subsection{Network Architectures}

The unimodal networks are generic MLP SNNs with increasingly compressed fully-connected layers moving forward through the network, and culminating in ten cumulative potential sum neurons on the output layer (Fig.~\ref{fig:unimodal_diagram}). These networks are implemented using a SNN Pytorch toolkit presented in \cite{sparch}.

Our multimodal structure has two individual networks for auditory and visual feature extraction. The direct spiking outputs of those networks are then concatenated and fed into a shared network which outputs a prediction based on the method of a cumulative sum of neuron potentials described in Section~\ref{sec:methods}. Our analysis consists of three different multimodal network architectures based on the depth of concatenation between the auditory and visual features described in Fig.~\ref{fig:multimodal_diagram}. The benefit of these architectural differences lies in how we focus our learning. Early concatenation increases the amount of data that the shared network sees in comparison to the unimodal networks, to increase performance. Late concatenation extracts high-level auditory and visual features before trying to learn the predictive task.

\subsection{McNemar $\chi^2$ Test}
\label{sec:mcnemar}

The null hypothesis for the McNemar test (with continuity correction) is that the performance of two models are the same. Rejecting the null hypothesis suggests two models disagree in different ways and could be considered to have statistically different performances. Combining this statistical test with accuracy scores allow us to evaluate two aspects. First, we can evaluate if our multimodal SNN models perform statistically better than our unimodal SNN models (Experiment 1 in Section~\ref{sec:results}). Second, we can determine if combining the visual and auditory branches at an early, middle, or late depth statistically influences the performance (Experiment 2 in Section~\ref{sec:results}).

\section{Results}
\label{sec:results}

\subsection{Model Performance}

\subsubsection{Experiment 1}
Displayed in Table~\ref{tab:accuracies}, our multimodal networks achieve an average test accuracy of 97.74\% at convergence, with our late-concatenation model scoring the highest accuracy of 98.43\%. The accuracy between each multimodal model are somewhat similar, all within about one percent. A McNemar test was performed between each model pairing using the test data with a desired hypothesis confidence of 95\% ($p$-value = 0.05). That is, a rejection of the null hypothesis indicates that the models have significantly different classification output with 95\% confidence. Experiment 1 in Table~\ref{tab:mcnemar} shows that our early branch concatenation multimodal SNN model performs statistically different than both the unimodal visual and unimodal auditory models ($p<0.001$). Therefore the null hypothesis is rejected, meaning our multimodal SNN model has a statistically better performance than each of the unimodal models. Both unimodal models perform similarly to one another according to a McNemar test ($p>0.1$). Our multimodal SNN models perform significantly better compared to both of our unimodal SNN models.

\begin{table}[t]
\caption{McNemar test table (using 1280 test instances) to analyze the differences in classifier performance for multimodal (MM) SNN models to unimodal (UM) SNN models and for combining branches at varying depths in MM SNN architecture.}
\begin{tabular}{lcccc}
             & \multicolumn{1}{c}{Model 1}   & \multicolumn{1}{c}{Model 2}   & \multicolumn{1}{c}{$p$-value}  &  \\
             \hline
Experiment 1 & \multicolumn{1}{c}{MM Early}  & \multicolumn{1}{c}{UM Visual} & \multicolumn{1}{c}{$p<0.01$}  &  \\ 
             & \multicolumn{1}{c}{MM Early}  & \multicolumn{1}{c}{UM Auditory}  & \multicolumn{1}{c}{$p<0.01$} &  \\ 
             & \multicolumn{1}{c}{UM Visual} & \multicolumn{1}{c}{UM auditory}  & \multicolumn{1}{c}{$p=0.241$}    &  \\
             &                               &                               &                              &  \\
Experiment 2 & MM Early                      & MM Middle                     & $p=0.473$                        &  \\
             & MM Early                      & MM Late                       & $p=1.00$                         &  \\
             & MM Late                       & MM Middle                      & $p=0.720$                        & 
\end{tabular}
\label{tab:mcnemar}
\end{table}
\subsubsection{Experiment 2}
Table~\ref{tab:mcnemar} shows that combining the visual and auditory branches at an early, middle, or late stage does not show a statistical difference in performance. Comparing our multimodal models reveals no difference in model output can be considered significant ($p>0.1$).
Therefore the null hypothesis is not rejected, meaning the performance of the multimodal models are the same regardless of our branch concatenation depths. This suggests that SNNs are adept at fusing information regardless of where the information is introduced. 

\section{Conclusion \& Future Work}
\label{sec:conclusion}

Our work examined the neuromorphic advantage of multimodal systems for a digit recognition task. We also find that information fusion at early, mid, and late depths results in similar performance for our multimodal SNN. This work can be extended to a more general classification task, such as image classification or emotion classification.

\bibliographystyle{IEEEbib}
\bibliography{refs}

\begin{thebibliography}{10}

\bibitem{mazzoni1991more}
Pietro Mazzoni, Richard~A Andersen, and Michael~I Jordan,
\newblock ``A more biologically plausible learning rule for neural networks.,''
\newblock {\em Proceedings of the National Academy of Sciences}, vol. 88, no.
  10, pp. 4433--4437, 1991.

\bibitem{eluyode2013comparative}
OS~Eluyode and Dipo~Theophilus Akomolafe,
\newblock ``Comparative study of biological and artificial neural networks,''
\newblock {\em European Journal of Applied Engineering and Scientific
  Research}, vol. 2, no. 1, pp. 36--46, 2013.

\bibitem{ghosh2009spiking}
Samanwoy Ghosh-Dastidar and Hojjat Adeli,
\newblock ``Spiking neural networks,''
\newblock {\em International journal of neural systems}, vol. 19, no. 04, pp.
  295--308, 2009.

\bibitem{tavanaei2019deep}
Amirhossein Tavanaei, Masoud Ghodrati, Saeed~Reza Kheradpisheh, Timothe
  Masquelier, and Anthony Maida,
\newblock ``Deep learning in spiking neural networks,''
\newblock {\em Neural networks}, vol. 111, pp. 47--63, 2019.

\bibitem{10.3389/fnsys.2022.845177}
Yuwei Wang and Yi~Zeng,
\newblock ``Multisensory concept learning framework based on spiking neural
  networks,''
\newblock {\em Frontiers in Systems Neuroscience}, vol. 16, 2022.

\bibitem{9746865}
Qianhui Liu, Dong Xing, Lang Feng, Huajin Tang, and Gang Pan,
\newblock ``Event-based multimodal spiking neural network with attention
  mechanism,''
\newblock in {\em ICASSP 2022 - 2022 IEEE International Conference on
  Acoustics, Speech and Signal Processing (ICASSP)}, May 2022, pp. 8922--8926.

\bibitem{10.3389/fnins.2018.00331}
Yujie Wu, Lei Deng, Guoqi Li, Jun Zhu, and Luping Shi,
\newblock ``Spatio-temporal backpropagation for training high-performance
  spiking neural networks,''
\newblock {\em Frontiers in Neuroscience}, vol. 12, 2018.

\bibitem{Kheradpisheh_2018}
Saeed~Reza Kheradpisheh, Mohammad Ganjtabesh, Simon~J. Thorpe, and
  Timoth{\'{e}}e Masquelier,
\newblock ``{STDP}-based spiking deep convolutional neural networks for object
  recognition,''
\newblock {\em Neural Networks}, vol. 99, pp. 56--67, mar 2018.

\bibitem{MIDDYA2022108580}
Asif~Iqbal Middya, Baibhav Nag, and Sarbani Roy,
\newblock ``Deep learning based multimodal emotion recognition using
  model-level fusion of audio--visual modalities,''
\newblock {\em Knowledge-Based Systems}, vol. 244, pp. 108580, 2022.

\bibitem{10.3389/fnins.2015.00437}
Garrick Orchard, Ajinkya Jayawant, Gregory~K. Cohen, and Nitish Thakor,
\newblock ``Converting static image datasets to spiking neuromorphic datasets
  using saccades,''
\newblock {\em Frontiers in Neuroscience}, vol. 9, 2015.

\bibitem{cramer_heidelberg_2020}
B.~Cramer, Y.~Stradmann, J.~Schemmel, and F.~Zenke,
\newblock ``The {Heidelberg} {Spiking} {Data} {Sets} for the {Systematic}
  {Evaluation} of {Spiking} {Neural} {Networks},''
\newblock {\em IEEE Transactions on Neural Networks and Learning Systems}, pp.
  1--14, 2020.

\bibitem{sparch}
Alexandre Bittar and Philip~N. Garner,
\newblock ``A surrogate gradient spiking baseline for speech command
  recognition,''
\newblock {\em Frontiers in Neuroscience}, vol. 16, 2022.

\bibitem{10.3389/fnins.2022.1079357}
Chengting Yu, Zheming Gu, Da~Li, Gaoang Wang, Aili Wang, and Erping Li,
\newblock ``Stsc-snn: Spatio-temporal synaptic connection with temporal
  convolution and attention for spiking neural networks,''
\newblock {\em Frontiers in Neuroscience}, vol. 16, 2022.

\bibitem{7966074}
Laxmi~R. Iyer and Arindam Basu,
\newblock ``Unsupervised learning of event-based image recordings using
  spike-timing-dependent plasticity,''
\newblock in {\em 2017 International Joint Conference on Neural Networks
  (IJCNN)}, 2017, pp. 1840--1846.

\bibitem{10.3389/fnins.2016.00508}
Jun~Haeng Lee, Tobi Delbruck, and Michael Pfeiffer,
\newblock ``Training deep spiking neural networks using backpropagation,''
\newblock {\em Frontiers in Neuroscience}, vol. 10, 2016.

\bibitem{kaiser}
Jacques Kaiser, Hesham Mostafa, and Emre Neftci,
\newblock ``Synaptic plasticity dynamics for deep continuous local learning
  (decolle),''
\newblock {\em Frontiers in Neuroscience}, vol. 14, 2020.

\end{thebibliography}

\end{document}